%% file: 0_main.tex
\PassOptionsToPackage{table,xcdraw}{xcolor}

\documentclass[sigconf, nonacm]{acmart}

\usepackage[table,xcdraw]{xcolor}
\usepackage{tabularx, booktabs}

\usepackage{array}
\usepackage{macros}
\AtBeginDocument{%
  \providecommand\BibTeX{{%
    \normalfont B\kern-0.5em{\scshape i\kern-0.25em b}\kern-0.8em\TeX}}}

\def\markup{1}

\if\markup1

\else

\newcommand{\st}[1]{}
\fi
\usepackage{soul}

\usepackage{hyperref}

\usepackage{xspace}
\usepackage[many]{tcolorbox}
\tcbuselibrary{listings}
\usepackage{wrapfig}
\usepackage{graphicx}
\usepackage{hyperref}
\usepackage{multirow}

\definecolor{decisioncolor}{RGB}{195, 132, 36}
\definecolor{dotcolortext}{RGB}{158, 94, 255}
\definecolor{dotcolorkeywords}{RGB}{73, 188, 73}
\definecolor{dotcolorgist}{RGB}{255, 170, 54}
\definecolor{dotcolorknowledge}{RGB}{245, 79, 79}

\newtcblisting{cognitivebox}{
  listing only,
  nobeforeafter,
  after={\xspace},
  hbox,
  tcbox raise base,
  fontupper=\ttfamily,
  colback=BLCognitive,
  colframe=BLCognitive,
  size=fbox,
  boxrule=0pt,        % No frame rule
  left=-1pt,           % No left padding
  right=-1pt,          % No right padding
  top=-4.5pt,            % No top padding
  bottom=-5pt,         % No bottom padding
}

\newtcblisting{foragingbox}{
  listing only,
  nobeforeafter,
  after={\xspace},
  hbox,
  tcbox raise base,
  colback=YLForaging,
  colframe=YLForaging,
  size=fbox,
  boxrule=0pt,        % No frame rule
  left=-1pt,           % No left padding
  right=-1pt,          % No right padding
  top=-4.5pt,            % No top padding
  bottom=-5pt,         % No bottom padding
}

\newtcolorbox{boxF}{
    colback = sub,
    enhanced,
    boxrule = 1.5pt, 
    colframe = white, % making the base for dash line
    borderline = {1.5pt}{0pt}{main, dashed} % add "dashed" for dashed line
}

\newtcolorbox{sensemakingbox}{
  listing only,
  nobeforeafter,
  after={\xspace},
  hbox,
  tcbox raise base,
  colback=white,        % Set background to white (transparent)
  colframe=white,  % Set frame color to YLForaging
  size=fbox,
  boxrule=0.5pt,        % Set the frame rule to 0.5pt (adjust as needed)
  boxsep=0pt,           % No padding inside the box
  left=.2pt,             % No left padding
  right=.2pt,            % No right padding
  top=.3pt,              % No top padding
  bottom=.2pt,           % No bottom padding
  enhanced,
  overlay={
    \draw[YLSensemaking, dashed, line width=1pt] (frame.north west)--(frame.north east)--(frame.south east)--(frame.south west)--cycle;
  },
}

\definecolor{Red500}{HTML}{EF4444}
\definecolor{LightGray}{HTML}{EBEBEB} 
\definecolor{DarkGray}{HTML}{4C4E52}
\definecolor{SH200}{HTML}{E3BDC5}
\definecolor{BLCognitive}{HTML}{C7DCFC}
\definecolor{YLForaging}{HTML}{F7F8C0}
\definecolor{YLSensemaking}{HTML}{CBCD1A}

\makeatletter
\def\@ACM@checkaffil{% Only warnings <<<<<<<<<<<<<<<<
    \if@ACM@instpresent\else
    \ClassWarningNoLine{\@classname}{No institution present for an affiliation}%
    \fi
    \if@ACM@citypresent\else
    \ClassWarningNoLine{\@classname}{No city present for an affiliation}%
    \fi
    \if@ACM@countrypresent\else
        \ClassWarningNoLine{\@classname}{No country present for an affiliation}%
    \fi
}
\makeatother

\begin{document}

%%
%% The title
\title[To Search or To Gen]{To Search or To Gen? Exploring the Synergy between Generative AI and Web Search in Programming}

%% The authors

\author{Ryan Yen}
\orcid{0001-8212-4100}

\affiliation{%
  \institution{University of Waterloo}
  \streetaddress{200 University Ave W}
  % \city{Waterloo}
  % \state{Ontario}
  % \country{Canada}
}
\email{r4yen@uwaterloo.ca}

\author{Nicole Sultanum}
\orcid{0001-8608-1427}
\affiliation{
  \institution{Tableau Research}
}
\email{nsultanum@tableau.com}

\author{Jian Zhao}
\orcid{0002-7761-6351}

\affiliation{%
  \institution{University of Waterloo}
  \streetaddress{200 University Ave W}
  % \city{Waterloo}
  % \state{Ontario}
  % \country{Canada}
}
% \authornote{Corresponding Author}
\email{jianzhao@uwaterloo.ca}

\newcommand{\sys}[0]{{{\it SearchNGen}}}
\newcommand{\baseline}[0]{\textit{Baseline}}

\begin{abstract}
The convergence of generative AI and web search is reshaping problem-solving for programmers. However, the lack of understanding regarding their interplay in the information-seeking process often leads programmers to perceive them as alternatives rather than complementary tools. To analyze this interaction and explore their synergy, we conducted an interview study with eight experienced programmers. Drawing from the results and literature, we have identified three major challenges and proposed three decision-making stages, each with its own relevant factors. Additionally, we present a comprehensive process model that captures programmers' interaction patterns. This model encompasses decision-making stages, the information-foraging loop, and cognitive activities during system interaction, offering a holistic framework to comprehend and optimize the use of these convergent tools in programming.
\end{abstract}

\keywords{generative AI, code generation, LLM, web search, information foraging, sensemaking}

%% A "teaser" image
%   \includegraphics[width=\textwidth]{Figs/teaser.png}
%   \caption{}
%   \Description{}

\maketitle

\input{1_intro}
\input{2_rw}

\input{3_design}

\input{4_system}

\input{7_discussion}

\input{9_conclusion}

% \begin{acks}
% To Robert, for the bagels and for explaining CMYK and color spaces.
% \end{acks}

\bibliographystyle{ACM-Reference-Format}
\bibliography{sample-base}

\appendix
\input{10_appendix}

\end{document}

%% file: 1_intro.tex
\section{Introduction}
Programmers often invest time in seeking and making sense of external information to tackle programming tasks~\cite{brandt2009two, liu2021reuse}. 
Traditionally, programmers frequently engage in web searches to resolve coding challenges, such as debugging. 
They rely on search engines to find relevant information, error messages, and solutions shared by others in the programming community~\cite{sadowski2015developers, hsieh2018exploratory}.
However, recent advancements in Large Language Models (LLMs) have introduced an alternative information-seeking approach, consisting of generating solutions via natural language prompts. 
With the ability to generate customized responses for various programming scenarios, programmers turn to generative AI for tasks such as producing boilerplate code or implementing external APIs.

This raises questions about the role and coexistence of these distinct information-seeking tools in programmer workflows.
Research comparing the two found that programmers often prefer generative AI over web searches when dealing with low-level code implementation due to their accessibility and adaptability~\cite{barke2023grounded}, but continue depending on web searches to explore diverse solutions and to acquire domain-specific terms that aid them in translating their vague goals into concrete prompts for generative AI~\cite{barke2023grounded, sarkar2022like, vaithilingam2022expectation}.
Another line of research seeks to integrate the functionality of these two methods by adopting retrieval-augmented generation (RAG)~\cite{lewis2020retrieval}. 
However, simply combining these two features contradicts programming practices, as programmers do not always accept the top search results. 
Instead, they rely on \textit{signals}, such as source credibility, to assess the suitability of the results~\cite{liu2021reuse, metzger2007making}.

While prior research has highlighted programmers' interest in combining the use of these two tools~\cite{vaithilingam2022expectation}, the lack of a clear understanding of their intersection often leads programmers to consider them as substitutes rather than synergistic.
Further, due to the inherent uncertainty and variance of both web search~\cite{brandt2009two} and generative AI~\cite{mozannar2022reading}, programmers often resort to opportunistically choosing between the two tools. 
To address these challenges, it is essential to \emph{understand the decisions that programmers make during the information-seeking process.}
This understanding will inform future designs that consider interactions with both tools.

We conducted interviews with eight programmers well versed in both web search and generative AI for information seeking, to learn about their common practices when using both tools for programming problem-solving.
Based on the results, we identified key challenges and three major decision stages, each with its own set of factors influencing the decisions. 
We synthesize these findings with existing literature and propose a process model by incorporating the interaction with generative AI into the classic information-foraging loop by Pirolli et al.~\cite{pirolli1999information}. 
This model outlines key stages of activities throughout the interaction, offering insights for future integrated designs that aim to assist programmers in effectively utilizing both web search and generative AI in their information-seeking processes.

%% file: 2_rw.tex
\section{Related Work}
To investigate the synergy between web search and generative AI, we conducted a review of programmers' information-seeking processes for both tools.

\subsection{Information Seeking and Knowledge Reusing from Web}
The dynamics of utilizing web resources are multifaceted in the field of web information seeking and knowledge reuse. 
Programmers actively seek relevant information across various domains as described in the Information Foraging Theory (IFT)~\cite{pirolli1999information}, such as code and search results~\cite{perez2014diagnosis, lawrance2007scents}. 
They not only gather information pertinent to their current issues~\cite{brandt2010example, hoffmann2007assieme, stylos2006mica} but also synthesize this information to create structured knowledge, aiding in their decision-making~\cite{kittur2013costs, kittur2014standing, liu2022crystalline}.
A prevailing challenge is comprehending the rationale behind previous programmers' decisions due to inadequate or outdated documentation~\cite{sillito2006questions, horvath2022using}, which underscores the importance of effective knowledge documentation and retrieval. 
In this context, knowledge reuse becomes significant, as it entails not just generating new knowledge but applying existing knowledge to problem-solving~\cite{hahn2018bento, davenport1996improving, li2019pumice, liu2019unakite, vermette2015cheatsheet}.
Building upon this previous literature, our study focuses on the intersection between information-seeking via the web and generative AI, particularly in terms of understanding, translating, and reusing knowledge from one tool to another.

\subsection{Generative AI-assisted Problem-Solving}
Similar to web searches, using generative AI for programming problem-solving also involves information-seeking and sense-making processes, yet with distinct workflows and challenges.
Programmers using generative AI in their problem-solving often encounter limitations within the linear question-answer paradigm, which hinder exploration and the ability to revisit previous responses~\cite{ross2023programmer, barke2023grounded, xu2022ide}. 
LLMs are also capable of generating plausible results for NL prompts that may not necessarily be aligned with the current usage scenario, which exacerbates friction~\cite{yao2023llm, liu2023wants, zamfirescu2023johnny, feng2023coprompt}.
For instance, writing \textit{``Scrape this web page with JavaScript''} can already generate a program without syntax error, suggesting the use of Node.js for web scraping. Nevertheless, programmers might also need to visualize the scraped data on a self-hosted website, which necessitates web scraping with a server-side rendering front-end JavaScript framework.
As a result, programmers might find themselves ensnared in a debugging rabbit hole~\cite{mozannar2022reading, zamfirescu2023johnny}, devoid of the guidance that web searches offer in terms of evaluating the credibility and suitability of particular answers~\cite{vaithilingam2022expectation}.

Prior work has also incorporated information retrieval techniques into the generation process of LLMs~\cite{lewis2020retrieval}, offering access to real-world and up-to-date information.
Our work believes programmers' active involvement in the information-seeking and sensemaking process is essential for them to iterate on prompts or search queries in their pursuit of the most suitable results.
Therefore, further investigation is warranted to explore the intersection, challenges, and requirements when interacting with both web search and generative AI.

%% file: 3_design.tex
\section{Interview Study}
To gain insights into the practices employed by programmers, challenges and factors in their decision-making, we conducted retrospective interviews with experienced programmers.

\subsection{Participants and Procedure}
We recruited eight participants ($5$ males, $3$ female; ages $24-29, M = 26.8, SD = 1.26$) through purposive sampling~\cite{etikan2016comparison}.
In our recruitment process, we sought participants experienced in programming and using LLM-driven code generation tools.
Eligibility screening involved a pre-test survey that assessed participants' self-reported programming experience on a 5-point scale [1: very inexperienced; 5: very experienced], years of programming experience, and self-reported familiarity with LLM-driven code generation tools (Pre-test survey in Appendix~\ref{appendix:prestudy}).
All recruited participants reported having more than four years of programming experience ($M = 5.43$ years, $SD = 1.12$) and were confident in their programming experiences (score $M = 4.29, SD = 1.08$), familiar with LLM-code generation tools (score $M = 4.41, SD = 0.37$), and regularly used the LLM-code generation tools ($M = 12$ times/week, $SD = 4.19$).

Participants were compensated with $20$ CAD for a $45$-minute interview session.
We first asked each participant to provide a minimum of three recent examples of their ChatGPT usage for programming problem-solving, including instances involving web search as part of the process, to encourage participants to reflect on their utilization of both web search and generative AI.
We then asked about challenges they faced when interacting with both tools, explored scenarios involving the combined usage, and inquired about their thought processes throughout the information-seeking process (Interview Questions in Appendix~\ref{appendix:interview}).

All interviews were audio-recorded and subsequently transcribed into written text.
We analyzed the interviews using thematic analysis~\cite{braun2012thematic}, employing both inductive and deductive approaches. 
After interviewing eight participants, the research team conducted the initial analysis collaboratively.
During this phase, an open coding approach~\cite{charmaz2006constructing} was employed to identify and categorize codes and themes. 
The focus was on challenges encountered, stages of decision-making, key factors influencing decisions, and types of knowledge extraction from results. 
Multiple iterations of discussions with the research team were conducted to resolve any discrepancies in coding and theme categorization.

\subsection{Collected Data}
All participants provided at least three example scenarios in which they employed both web search and generative AI for information-seeking and problem-solving. 
In the majority of scenarios (26 out of 28), participants engaged in more than two rounds of iterations involving both web search and prompting, with two cases involving only one web search session and one round of prompting.

Regarding the tasks participants undertook, we identified 21 of them as open-ended tasks where participants did not require specific approaches to solve them. Among these, 9 pertained to exploratory data analysis and modelling tasks, 5 involved front-end development, 4 related to data mining and web scraping, and the remaining 3 concerned server-related tasks.

% This suggests a reliance on internalized strategies to process the knowledge acquired during the process.

% Thus, scaffolding is needed to decompose the process for programmers to make more informed decisions while preserving the provenance of their thought process.

% \subsubsection{Roles of Web Search and GenAI in Information-Foraging Process}

% \subsubsection{Key Facets Affecting Decision-Making}

% \subsubsection{Challenges and Needs for Leveraging Both Mediums}

%% file: 4_system.tex
\section{Challenges}
Results from the interview study indicate that programmers require assistance in making informed decisions about which tool to use (C1), determining which results to extract and apply (C2), and translating results into concrete search queries or prompts for subsequent iterations (C3).

\subsection{C1 - Lack of Guidance to Determine Tool Selection and Integration}
In the majority of scenarios presented by participants, either web search or generative AI was seen as a fallback to the other when each failed. 
For example, P6 mentioned, \qt{I would try using web search when ChatGPT kept giving me the wrong answer.}
There were also several scenarios where combining both tools produced the most favourable outcomes, as \pqt{they each possessed their strengths.}{P2}. 
Additionally, participants did not report explicit metrics to determine which tool should be used next, stating they \pqt{cannot anticipate how the results will appear}{P1} or \pqt{what knowledge I [they] will gain}{P3} in the current round.
The choice of what approach to try next still relies on trial and error. 
Therefore, it is relevant to comprehend the factors that influence the choice and offer programmers guidance on which tool to use in various circumstances.

\begin{figure*}[th]
    \centering
    \includegraphics[width=.9\linewidth]{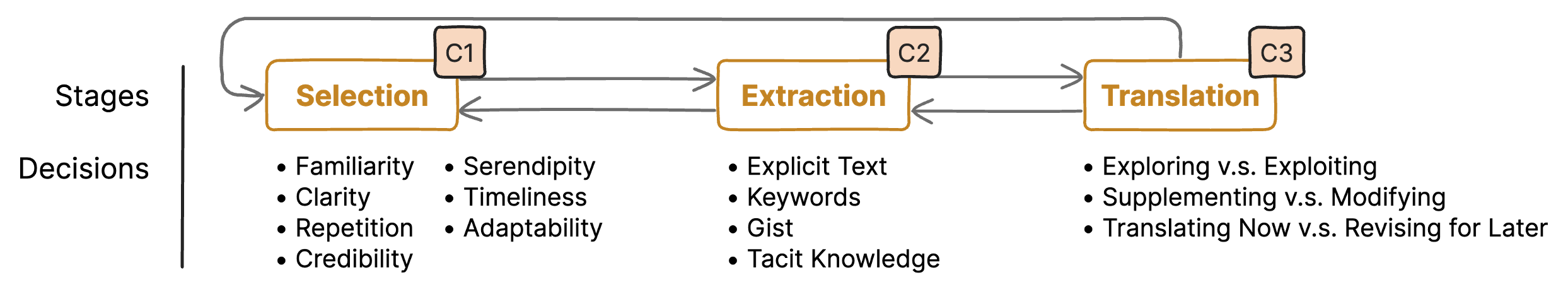}
    \caption{Three decision-making stages with key factors influencing these decisions and the challenges associated with them.}
    \label{fig:decision-stages}
\end{figure*}

\subsection{C2 - No Scaffolding to Extract Information from Results}
Most participants have developed their own strategies for understanding and extracting information gained from either generated results or web search results.
However, the majority of participants (7 out of 8) \emph{did not mention specific metrics to evaluate the appropriateness of the results}.
For instance, P2 stated that the appropriateness of generated results was mostly based on \qt{experience} and \qt{when the results are similar to what I [they] expected}.
In most scenarios, participants were unable to explicitly express the sources of the knowledge they extracted from. 
For instance, P4 mentioned, \qt{I did write a new prompt based on the knowledge I gained from web pages I visited, but I could not tell exactly where it is from.}
This highlights the need for provenance tracking for externalizing programmers' information-seeking process, consistent with prior research~\cite{palani2021conotate}.

\subsection{C3 - Difficulties in Adapting Results between Search and Generative AI}
All participants reported difficulties in switching between the two tools due to the lack of understanding regarding \pqt{the exact differences}{P5} between them.  
Participants often struggled with translating results from one format to another to derive \pqt{the best outcome from both inputs.}{P5}
Some (3 out of 8) also explicitly expressed a desire for a more seamless integration of both tools.

\section{Decision-Making Stages}
To summarize the activities of programmers during the information-seeking process, we outline three decision stages that programmers encounter, each associated with the challenges mentioned above (see Fig.~\ref{fig:decision-stages}):
\begin{enumerate}
    \item The 
    \textcolor{decisioncolor}{\textit{Selection}} stage, during which programmers choose between the two available tools (C1).
    \item In the \textcolor{decisioncolor}{\textit{Extraction}} stage, programmers assess whether to utilize implicit or explicit knowledge (C2).
    \item During the \textcolor{decisioncolor}{\textit{Translation}} stage, programmers transform their internal knowledge into natural language queries or prompts (C3).
\end{enumerate}

\subsection{Selection Stage}
We identified seven major factors that influence programmers' decisions on \textit{which tools to use in each iteration}. 

\subsubsection{Familiarity with the Domain}
Participants use web searches to familiarize themselves with how to construct their prompts, particularly when they are not familiar with the domain.
In such scenarios, participants understand that the results generated by AI may not provide the answer, and they \pqt{fear that I [they] do not possess the knowledge to verify correctness.}{P1}.
Conversely, participants tended to employ generative AI when they were more \textit{familiar} with the task at hand, such as using APIs they may have forgotten or implementing detailed algorithms.

\subsubsection{Clarity of Goals}
Participants generally turned to web searches when their goals were unclear and less defined. 
In such situations, they struggle to \pqt{determine which keywords are appropriate}{P6} and which approaches to take. The diversity of web search results offers them a better understanding of potential solutions.
Once the problem becomes more well-defined, participants prefer to use generative AI, as their primary objective at that point is to \pqt{find the most aligned [generated] results.}{P1}.
Another condition that leads participants to opt for generative AI over web search is when they believe that the problem is well-defined and has been thoroughly discussed in the past. 
P5 explained, \qt{I believe this problem has enough solutions being trained as data to the [AI] model.}

\subsubsection{Repetition of Results}
Participants often find themselves trapped in a debugging cycle where they continuously iterate on the prompt and repetitively receive incorrect results. 
Previous research has reported this issue when users repeatedly receive results that do not align with their intentions, causing them to struggle to find the right terms to prompt the LLM.
The majority of participants (6 out of 8) mentioned that they would fallback to web searches when they realized they were \pqt{stuck in the loop.}{P2} P6 elaborated, \qt{I will still try a few times if the [AI-generated] results do not match before turning to the web search.}

\subsubsection{Credibility and Diversity}
Participants believe that by glancing through the appropriateness of multiple search results, they can easily \pqt{identify the keywords to use for prompts}{P8} and determine the \pqt{overall approach they should take}{P3} to solve the problem, which echos prior research~\cite{eysenbach2002consumers, metzger2007making}.
Four participants mentioned that they would use web searches to validate the correctness of the generated results before applying them. 
They suggested that web searches provide \pqt{more detailed explanations from different aspects}{P3} and provide additional signals for assessing credibility, such as upvotes on StackOverflow.

\subsubsection{Serendipity and Luck}
While we observed that most circumstances begin with a web search when the problem is vague, we also noticed several scenarios where participants opt for generative AI in the hope that it might opportunistically provide the final result.
This finding echoes previous research that uncovered similar behaviours~\cite{zamfirescu2023johnny, nam2023ide}, similar to opportunistic programming~\cite{brandt2008opportunistic}.

\subsubsection{Up-to-dateness of Information}
% how up to date is the results
The timeliness of information is usually considered a crucial aspect of resource credibility~\cite{metzger2007making, tate2009web, liu2021reuse}. 
Participants reported that they tend to opt for web searches when they require the most up-to-date results, particularly when they seek the latest documentation for packages or libraries.
However, participants pointed out that the overall approach they follow does not necessarily have to be up to date; instead, only the low-level code implementation requires currency. 
For instance, when P6 was working on creating a responsive design, the high-level strategy remained consistent, but the low-level code required updates to align with current browser capabilities (e.g., specific CSS properties for responsiveness).

\subsubsection{Customizability and Adaptability}
When adapting solutions back into their programs, all participants relied on generative AI. 
They preferred this approach because of the customizability and adaptability of the generated results. 
P5 explained that \qt{GPT kind of combines solutions for you,} and P7 mentioned that \qt{I do not have to change the parameters or variables.}
However, two participants voiced concerns about overconfidence, as they occasionally bypassed the validation phases and overlooked inaccuracies.

\subsection{Information Extraction Stage}
At this stage, programmers must discern \textit{which results are useful for the next iteration}. We identified two relevant factors.

\subsubsection{Role of Early Iterations}
All participants reported that the results from the first few iterations are not worth reading in detail, especially when the problem is vague or relatively complex. 
The primary goal of these initial iterations is to \pqt{narrow down the scope [of the solution]}{P6} and \pqt{define the problem domain.}{P1}. 
Most participants (7 out of 8) also mentioned that these initial rounds serve to understand whether the search query or prompt can provide results in the \pqt{right direction.}{P7}. 
Thus, participants need to quickly skim through either the search results or the generated content to identify any misalignment with their intentions.
For example, P2 asked the generative AI to visualize a dataset in a bubble chart with a prompt like \qt{Scatter plot with some dots are larger [...].} The results showed a scatter plot where the marker size changed. 
P2 skimmed through the visualization part in the step-by-step tutorial and realized that implementing a scatter plot might be too complex.
Participants then had to decide whether to switch to another tool or continue iterating to obtain a more aligned solution.

\subsubsection{Extract Text, Keywords, Gist, or Tacit Knowledge}
We identified four major elements that programmers reused and carried forward to the next iteration through iterative open coding.
At the most concrete level, participants often directly copy-pasted extracted $\textcolor{dotcolortext}{\bullet \textit{text}}$ from results either into the search query or as the context of the prompt. 
This was especially common when search results were lengthy, and participants preferred not to organize them themselves.
When iterating on the prompt or search query itself, most participants derived $\textcolor{dotcolorkeywords}{\bullet \textit{keywords}}$  from previous iterations and used them to guide the direction of the next solutions.
In some scenarios, participants simply translated the $\textcolor{dotcolorgist}{\bullet \textit{gist}}$ of their articulation to the next round without explicitly copying from specific parts of the results. 
This approach was evidenced when participants started another round right after scrolling through search results without clicking on any pages.
In the most abstract levels, participants applied implicit $\textcolor{dotcolorknowledge}{\bullet \textit{knowledge}}$ gained from the results. However, they did not always apply it directly to the next round; instead, it was sometimes more \pqt{useful when verifying the next results.}{P8}.

\subsection{Knowledge Translation Stage}
In the final stage of each iteration, programmers must decide \textit{how to translate their knowledge into text}, whether it be a search query or a prompt for generative AI. 
We have identified three quandaries that programmers should weigh as they trade off various considerations.

\subsubsection{Exploring vs. Exploiting}
The most common dilemma is the decision of whether to persist with iterations on the current topic or explore a wider range of topics. 
This finding aligns with the exploration and acceleration modes discovered by Barke et al~\cite{barke2023grounded}.
Participants either rewrote the entire prompt or query to ask different questions or refined the existing prompt or query to dive deeper into the same domain.
The challenge arises when programmers are uncertain \pqt{if the information is sufficient}{P7} for either generative AI or crafting web search queries. 
Programmers might proceed to the next exploration without fully grasping the solution. Consequently, they encounter difficulties in tracing back to search or prompt histories without analytical provenance~\cite{xu2015analytic, nguyen2016sensemap, north2011analytic}.

\subsubsection{Supplementing vs. Modifying}
Participants have the option to either add results into the prompt or query as context or directly modify the original prompt or query. 
The former is more common when programmers have clearer intentions in mind, while the latter is employed when they are still in the phase of adjusting the direction of information seeking.
For example, P4 sought information about implementing an API in the Next.js framework using generative AI. Initially, P4 tended to modify the prompt to understand where to implement the code and the differences between code versions. After deciding on approaches, P4 then pasted all the documentation and tutorials as context for AI to generate results.

\subsubsection{Translating Now vs. Reserving for Later}
Similar to previous research~\cite{mozannar2022reading}, which suggests that programmers often set aside some generated results for later use, our findings indicate that several participants (5 out of 8) may not immediately apply the knowledge acquired in one round to the subsequent round. 
Rather than a \textit{linear chain} of knowledge, this process resembles a more intricate \textit{tree diagram}, with various branches and connections as participants navigate and adapt their strategies. 
Participants can revisit and apply acquired knowledge later after exploring different branches. For instance, they might explore one branch, then switch to another, and eventually, go back to the first branch to combine insights for a more comprehensive solution.

\begin{figure*}[th]
    \centering
    \includegraphics[width=.9\linewidth]{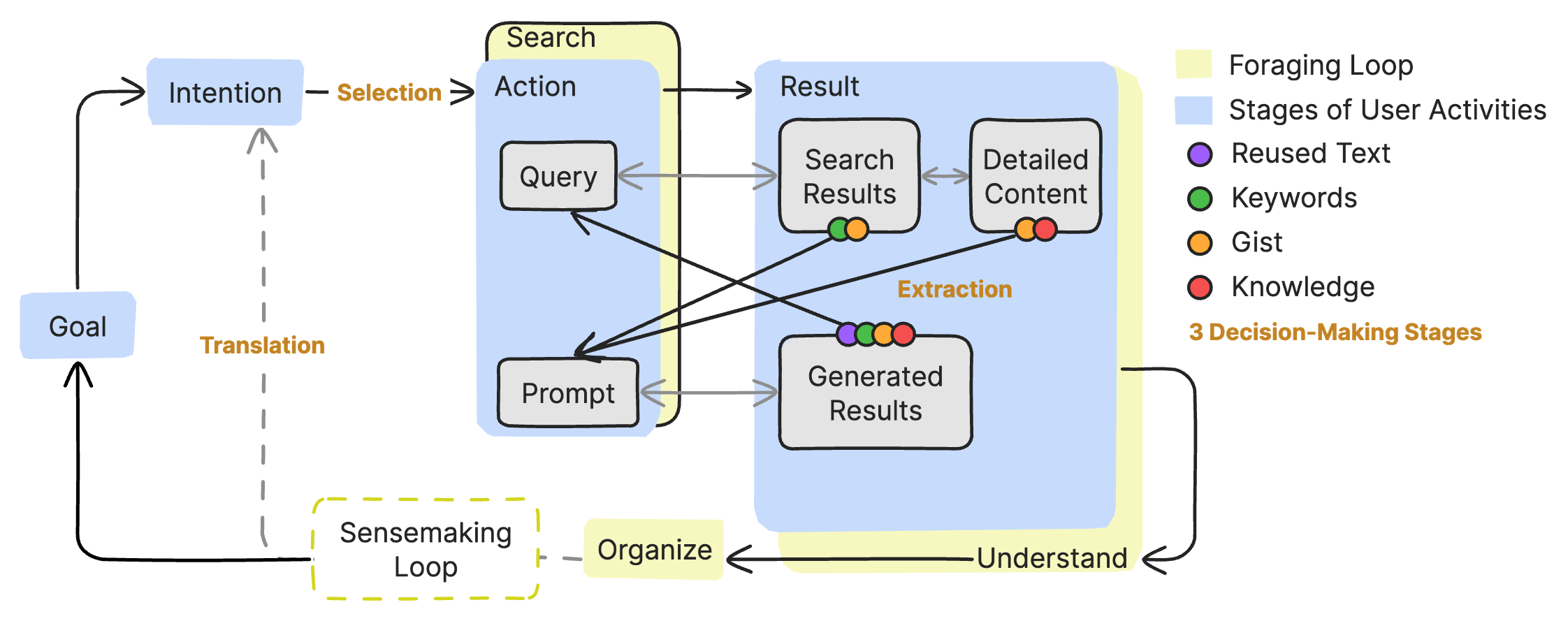}
    \caption{A process model describing interactions between programmers and search/generative AI.}
    \label{fig:model}
\end{figure*}

%% file: 7_discussion.tex
\section{Towards a Process Model
for Programmer-Search/Generative AI Interaction}

When interacting with both web search and generative AI, programmers face challenges at each decision stage, particularly in selecting the tool to use, determining which information to extract, and translating information into concrete text. 
Prior research has described the stages of information foraging~\cite{rachatasumrit2021forsense, suh2023sensecape, pirolli2005sensemaking}, offered frameworks for determining information appropriateness~\cite{liu2021reuse, hsieh2018exploratory}, and explored cognitive models for translating vague goals into natural language prompts in AI-driven systems~\cite{subramonyam2023bridging, yen2023coladder, norman1986cognitive}. 
However, without a clear understanding of the synergy between these two tools, current designs often focus on one or the other individually instead of considering them as a whole. 
While web search has been integrated further into generative AI~\cite{lewis2020retrieval}, current tools have not fully taken into account the interaction among information foraging loop and the value of the human cognitive thought process.

To help guide future design, we summarize our findings in the form of a \textit{process model} that enables parallel interactions with web search and generative AI. 
This model unifies our insights and those of prior studies, including principles of the sensemaking foraging loop~\cite{pirolli2005sensemaking} and Norman's seven stages of activities~\cite{norman1986cognitive} (see Fig.~\ref{fig:model}) to depict decision-making iterations.
The flow begins with a 
\begin{cognitivebox}
Goal
\end{cognitivebox}
, where programmers identify their problems and tasks.
In the pursuit of goals, programmers then articulate
\begin{cognitivebox}
Intentions
\end{cognitivebox}
for addressing them. 
For instance, when aiming to classify the Iris dataset,
% ~\footnote{\fixme{footnote or reference needed}}
programmers dissect the problem into selecting an appropriate model and outlining the training and evaluation procedures. 
They then either leverage web searches or generative AI systems as
\begin{cognitivebox}
Actions
\end{cognitivebox} 
. We termed this decision point as the \textcolor{decisioncolor}{\textit{Selection}} stage, which involves choosing the most suitable tool for the given context.

Upon executing their actions, programmers receive a list of search results or AI-generated content. 
Akin to the foraging loop's
\begin{foragingbox}
Understanding
\end{foragingbox}
step, they analyze and interpret these results, identifying essential information for \textcolor{decisioncolor}{\textit{Extraction}}. Continuing with our example, a programmer might encounter various model suggestions in their search results, select the keyword \qt{logistic regression}, and delve deeper into this specific approach. 
The \textcolor{decisioncolor}{\textit{Extraction}} stage involves distilling information from the results, encompassing explicit $\textcolor{dotcolortext}{\bullet \textit{text}}$, relevant $\textcolor{dotcolorkeywords}{\bullet \textit{keywords}}$, the $\textcolor{dotcolorgist}{\bullet \textit{gist}}$ from the overall results, and abstract $\textcolor{dotcolorknowledge}{\bullet \textit{knowledge}}$ gained during the process.

Programmers then 
\begin{foragingbox}
Organize
\end{foragingbox}
this acquired knowledge, preparing it for the \textcolor{decisioncolor}{\textit{Translation}} stage. 
This final decision point involves reformulating the information into a new search query or AI prompt. For example, they might integrate steps from a web tutorial with code examples to guide generative AI.
These activities among the
\begin{cognitivebox}
Gulf of Evaluation
\end{cognitivebox}
are essentials to determine the next step in the iteration. 
It can lead back to the 
\begin{cognitivebox}
Goal
\end{cognitivebox}
stage, particularly when the initial objective is broad and requires refinement, or return to the
\begin{cognitivebox}
Intention
\end{cognitivebox}
stage, where programmers refine their mental models for problem-solving. 
This cyclical process dynamically adapts to the evolving needs and understanding of the programmer, facilitating effective use of both web search and generative AI tools.

\subsection{Incorporating the Sensemaking Loop}
While this study did not directly investigate the
\begin{sensemakingbox}
sensemaking 
\end{sensemakingbox}
process, it is noteworthy that three participants mentioned the importance of note-taking to organize their collected information. 
This finding aligns with prior research indicating that programmers engage in a sensemaking loop before embarking on the next iteration of information foraging~\cite{pirolli2005sensemaking, suh2023sensecape, nguyen2016sensemap, rachatasumrit2021forsense}.
Incorporating this sensemaking process into the process model typically occurs during the evaluation process, starting from the
\begin{cognitivebox}
results
\end{cognitivebox}
.
Programmers articulate the knowledge they have internalized before making adjustments to their 
\begin{cognitivebox}
intentions
\end{cognitivebox}
or
\begin{cognitivebox}
goal
\end{cognitivebox}
. 
Future research could explore how programmers externalize their curated knowledge to generate results with context. 
Additionally, investigating how sensemaking influences the proposed decision stages, with the potential to enhance the translation stage, would be valuable.

%% file: 9_conclusion.tex
\section{Conclusion}
In this paper, we investigated the intersection of web search and generative AI in programming, uncovering key challenges and decision-making stages. 
Our findings from interviews with experienced programmers reveal a nuanced relationship between these tools, highlighting the need for a synergistic approach rather than treating them as alternatives. 
We then propose a process model that integrates web search and generative AI into programming workflows, emphasizing the importance of understanding, extracting, and translating information across tools.

%% file: 10_appendix.tex
\section{Survey and Interview Questions}

\section*{Eligibility Screening Survey}
\label{appendix:prestudy}

\begin{enumerate}
  \item \textbf{How confident are you in your overall programming experience?} 
    \begin{itemize}[label={}]
      \item[1:] Very Inexperienced
      \item[2:] Inexperienced
      \item[3:] Moderately Experienced
      \item[4:] Experienced
      \item[5:] Very Experienced
    \end{itemize}

  \item \textbf{How many years of programming experience do you have?} \small years

  \item \textbf{How familiar are you with AI code generation tools (e.g., GitHub Copilot, ChatGPT)?}
    \begin{itemize}[label={}]
      \item[1:] Not Familiar
      \item[2:] Slightly Familiar
      \item[3:] Moderately Familiar
      \item[4:] Familiar
      \item[5:] Very Familiar
    \end{itemize}

  \item \textbf{Over the past few weeks, how often did you typically employ AI code generation tools such as OpenAI's Codex, GitHub Copilot, or ChatGPT for your programming tasks?} \small(e.g., times per week)
\end{enumerate}

\subsection{Semi-Structured Interview Questions}
\label{appendix:interview}

\subsection*{Programming Workflow and Tool Integration}
\begin{enumerate}
  \item \textbf{Programming Workflow Integration:} 
  Can you describe your typical programming workflow, particularly emphasizing how you utilize code synthesis tools like Copilot and web search in this process?

  \item \textbf{Decision Making between GPT and Web Search:} 
  How do you decide when to use tools like GPT and when to resort to web search during your coding process? Could you provide a specific example illustrating this decision-making process?
\end{enumerate}

\subsection*{Information Seeking and Evaluation in Programming}
\begin{enumerate}
  \item \textbf{Information Requirements:}
  When you begin looking for information during programming, what specific types of information are you usually seeking? (e.g., syntax clarification, algorithmic approaches, best practices)

  \item \textbf{Assessment of Information Quality:} 
  What criteria do you use to determine whether a search or generated result is good enough for your needs? What factors are important to you?

  \item \textbf{Determining Importance of Information:} 
  What kind of information do you consider as most important from the generated or search results?

  \item \textbf{Synthesizing Information from Multiple Sources:} 
  Can you describe how you synthesize or combine information from different sources (like Copilot, web search, forums)? How do you resolve conflicts or discrepancies in information?

  \item \textbf{Long-term Information Retention:} 
  When you find particularly valuable information, how do you ensure its retention for future use? Do you have a system for organizing or bookmarking useful resources?
\end{enumerate}

\subsection*{Challenges and Limitations of Both Tools}
\begin{enumerate}[resume]
  \item \textbf{Challenges and Limitations:} 
  Could you discuss some challenges or limitations you've encountered with both tools? How does the other tool help in overcoming these challenges?

  \item \textbf{Web Search Efficacy:} 
  Can you provide an example where web search helped you gain a better understanding of a programming concept or language feature that tools like Copilot alone couldn't provide?
  
  \item \textbf{Future of GPT and Web Search:} 
  In your opinion, do you think the advancement of technologies like GPT-4 with internet and web scraping access could eventually replace traditional web search for programming-related queries?
\end{enumerate}